\documentstyle[aps,preprint,epsf]{revtex}
\begin{document}

\centerline{\today}
\vspace{0.5cm}
\centerline{\Large\bf Vortex matter in mesoscopic superconductors}
\vspace{0.5cm}
\centerline{J. J. Palacios}
\centerline{\em Departamento de F\'{\i}sica Te\'orica de la Materia Condensada} 
\centerline{\em Universidad Aut\'onoma de Madrid, Cantoblanco, 
Madrid 28049, Spain.}

\begin{abstract}
Superconducting mesoscopic devices in magnetic fields present novel
properties which can only be accounted for by both the quantum
confinement of the Cooper pairs and by the
interaction between the magnetic-field-induced vortices. Sub-micrometer disks, 
much the same as their semiconductor counterparts known as quantum dots,
are being subject to experimental investigation
by measuring their conducting properties and, more recently, their
magnetization by using state-of-the-art ballistic Hall magnetometry.
In this work I review the main results obtained in these two types of 
experiments as well as the current theoretical developments
which are contributing to our understanding of the superconducting
condensate in these systems.  
\end{abstract}


{\em Correspondence}: Juan Jos\'e Palacios

{\em Address}: Departamento de F\'{\i}sica Te\'orica de la Materia Condensada,

Universidad Aut\'onoma de Madrid, Cantoblanco, Madrid 28049, Spain

{\em Fax}: 34 91 397 4950

{\em E-mail}: palacios@kim.fmc.uam.es

\newpage

\section{Introduction}

The physics of mesoscopic systems 
has reached its maturity within the semiconductor community over the 
past few years\cite{semi_review}. 
A similar revolution is nowadays taking place in 
superconductivity where superconducting device miniaturization is
moving forward at a fast pace\cite{Tinkham}. In both communities
the concept of mesoscopics is directly 
linked to that of low-dimensionality. A system is called low-dimensional
when at least one of the dimensions in which its carriers live
becomes comparable to some relevant length scale 
intrinsically associated to bulk properties of those carriers. 
In semiconductors the most
relevant length scale one can think of is the Fermi wavelength
$\lambda_{\rm F}$. In superconductors there are two fundamental length scales.
One is the penetration length $\lambda$ over which an externally applied 
magnetic
field $H$ is screened into the superconducting condensate. The other one is
the coherence length $\xi$ which, in a simple-minded picture, can be thought
of as the size of the Cooper pairs and which limits the distance
over which the superconducting order parameter can vary appreciably. 
When one of the dimensions of a semiconductor becomes
comparable to $\lambda_{\rm F}$, the other two being much larger,
a two-dimensional electron gas (2DEG) is formed\cite{semi_review}. 
Similarly, typical superconducting thin films\cite{Tinkham} can have  
a width comparable to either $\lambda$ or $\xi$, 
or to both of them. If, in semiconductors, two out of the three
dimensions become comparable to $\lambda_{\rm F}$ one forms a quantum
wire\cite{semi_review}. A thin strip\cite{jjp_strip} can be thought of
as the superconducting counterpart. When the three
dimensions become comparable to either $\lambda_{\rm F}$ in semiconductors and
to $\lambda$ and/or $\xi$ in supercondutors one speaks of
'zero-dimensional' systems called quantum
dots\cite{Ashoori,Kouwenhoven,jjp_dot} and thin 
disks\cite{Moshchalkov,Geim,jjp_disk}, 
respectively. When the dimensions of the superconducting device become much
smaller than $\xi$ (as can be the case in very small metallic grains) 
the very notion of superconductivity needs to be revised\cite{grains}.
The above classification, far from being unique and rigorous, 
only pretends to serve as a guide for the non-specialized reader.

It has been known for a long time that, in addition to the
well-understood type-I
superconductors where  $\kappa = \lambda/\xi < 1/\sqrt{2}$,
superconductors with $\kappa > 1/\sqrt{2}$
are also possible (high-$T_c$ materials are the most exotic
example of them\cite{Tinkham}). This seemingly inocent relation between the two
fundamental lengths gives rise to type-II
superconductivity which is characterized by the appearance of 
vortices in the superconducting condensate at values of $H$ lying    
between a lower critical field
$H_{\rm c1}$ and an upper critical field $H_{\rm c2}$. An isolated vortex
extends over a distance of the order of $\lambda$ and exhibits a
normal core of radius of the order of $\xi$\cite{Tinkham}. 
A macroscopic number of
vortices generically form a triangular lattice\cite{Abrikosov} which presents
two well-characterized regimes or states: (i) 
a dilute vortex state (dVS) at low fields  in which $\lambda$
governs the inter-vortex distance and (ii) a dense vortex 
state (DVS) at high fields in which the fundamental length scale 
is given by $\xi$ since the vortices are closely packed and their
cores overlap strongly. In
extreme type-II superconductors, characterized by $\kappa \gg
1$, the crossover from the dVS to the DVS takes place at rather low
fields $H \approx 0.3H_{\rm c2}$\cite{Brandt} 
which gives an idea of the relative importance of both regimes.

The dVS has been thoroughly studied over the years in all possible
low-dimensional systems, both theoretical and experimentally. Many
experimental techniques are sensitive to the vortex lattice in this regime,
mainly due to the strong spatial modulation of the associated 
magnetic induction\cite{Brandt}. 
On the theory part, vortices in the dVS can be treated as 
classical, logarithmically-interacting, point-like particles which renders the 
calculation of the vortex structure free energy rather feasible in many 
different geometries\cite{dVS,Buzdin_disk,Venegas}. By contrast, 
the structure of the DVS in nanostructures is difficult to unveil, partially 
due to the 'invisibility' of this state to the usual experimental techniques 
and to the difficulty in minimizing the free energy associated to this
highly compacted vortex state when interfaces are present. Motivated
by very recent advances in transport\cite{Moshchalkov} and magnetometry
techniques\cite{Geim}, which we review in  Sec. II, 
we focus in this work on the theoretical study of the structural and magnetic
properties of the DVS in superconducting thin disks or quantum dots. 
In Sec. III we show how to do this within the traditional
framework of the Ginzburg-Landau functional in a rather analytical
fashion.  Finally, in Sec. IV we present our conclusions.

\section{Two relevant experiments in superconducting mesoscopic disks}

One the most remarkable achievements in low-dimensional semiconductor
physics has been the fabrication of the quantum dot or 
single-electron transistor where the electrons travel through the system
one at a time and where the number of them present in the device 
can be tuned at will even down to a single electron\cite{Ashoori,Kouwenhoven}. 
These systems have been also given the name of 
artificial atoms since their generic properties are determined 
by both the quantum confinement and the interaction among the 
electrons much the same
as in real atoms. Many analogies can be drawn between these artificial atoms
and superconducting mesoscopic disks in perpendicular magnetic fields.
Cooper pairs in disks can also experience the effects of quantum
confinement and this becomes visible in the superconducting-normal phase
boundary\cite{Moshchalkov}. As far as the interaction is concerned
the role of the electron is now played to some extent, not by the
Cooper pair, but by another fundamental entity: The vortex. 
When the dimensions of the disk are comparable to $\lambda$ in the dVS
or to $\xi$ in the DVS only few vortices can coexist in the disk.
In contrast to the usual triangular arrangement in bulk, complex and unique 
vortex structures are expected to occur due to the competition between
the geometrical confinement and the vortex-vortex interaction.

Transport experiments contributed in a decisive way to unveil the
electronic structure of artificial atoms\cite{Kouwenhoven}. Similarly,
conductance measurements\cite{Moshchalkov} in
individual mesoscopic disks gave us the first experimental evidence  of
the structure of the order parameter in these systems. Based on the
onset of the disk resistance, oscillations of the 
critical temperature $T_c$ as a function of the external magnetic field were
measured and 
correctly accounted for by the quantization of the angular momentum $L$ 
of the Cooper pair wavefunction. In more traditional words, they were
observing  transitions between
giant vortex states with a different number $L$ of fluxoids. These
oscillations were clearly reminiscent of those observed in the famous
Little-Parks experiment\cite{Little_Parks}. 

Transport measurements are necessarily bound to give only information
about the critical phase boundaries since they are based on 
measuring the resistance of the non-fully superconducting state of the disk.
Hall magnetometry\cite{Geim}, on the other hand, is revealing itself as a
powerful tool for obtaining information of the order parameter
away from the supercondutor-normal phase boundary.  Thin superconducting
Al disks are placed on top of a series of typical Hall probes created in a
2DEG. In the ballistic regime the Hall resistance is directly proportional 
to the average value of the magnetic field through the junction which,
in turn, is determined by the magnetic state of the disk lying right
above the junction. Magnetization ($M$) curves for different disk 
sizes can thus be obtained to
exhibit a variety of unexpected phase transitions. For very small disks
a first-order transition at a critical $H_{\rm c}$, which is what one expects
for a type-I superconductor as Al, is
absent. For larger sizes this transition appears, but, on
increasing further the disk radius the familiar second-order phase
transition for a type-II superconductor, namely, an steady decrease of the 
magnetization beyond a certain critical $H_{\rm c1}$,
becomes notorious. On top of this steady
decrease, the magnetization exhibits many jumps
corresponding to first-order transitions which 
present an irregular and decreasing height as a function of $H$.
All this seems to  indicate that the quantum confinement is also
patent in this experiment and that the Al disks are
behaving like type-II superconductors rather than 
type-I, possibly due to the expected enhancement of the 
effective magnetic penetration length in such a geometry.

In the next section we show how many of these unexpected 
features can be obtained and explained within
the traditional, phenomenological  Ginzburg-Landau theory.

\section{Ginzburg-Landau functional approach to the DVS in
superconducting mesoscopic disks}

The theoretical analysis of the peculiar
supercondutor-normal phase boundary measured in Ref.\
\onlinecite{Moshchalkov} does not present any difficulty 
since it simply implies solving the linearized Ginzburg-Landau
equations\cite{Moshchalkov,Benoist,Buzdin,Bezryadin,Moshchalkov2}.
The theoretical efforts to calculate the properties of the 
superconducting condensate in a disk well into the superconducting 
phase have been mostly restricted to the dVS\cite{Buzdin_disk,Venegas}.
As far as the DVS is concerned, recent work is being done in the direction
of solving numerically the Ginzburg-Landau equations either under the 
simplifying assumption
of an order parameter with a well-defined $L$\cite{Moshchalkov2,Peeters}
or without any symmetry restrictions\cite{Geim,newpeeters1,newpeeters2}. 
For type-II disks, the assumption of an order parameter with axial
symmetry (well-defined $L$) does not hold. More
precisely, it is only expected to hold in the Meissner state, which is
associated to an $L=0$ order parameter, 
and above $H_{\rm c2}$\cite{note} where, close to the surface of the disk,
the superconductivity can survive up to a
higher critical field $H_{\rm c3}$\cite{Moshchalkov,Benoist,Buzdin,Bezryadin}. 
This surface superconductivity is referred to as a condensed state of
vortices or giant
vortex\cite{Moshchalkov2,Peeters,newpeeters1,newpeeters2}. 
For $H_{\rm c1}<H< H_{\rm c2}$ it was argued\cite{Moshchalkov2} and recently 
shown in numerical simulations\cite{Geim,newpeeters2} 
that the order parameter can form complex structures of single-fluxoid 
vortices, i.e., a ``budding" Abrikosov lattice. 
From our more analytic standpoint the appearance of these structures
necessarily implies an order parameter without a well-defined angular 
momentum $L$. Recent work in which this condition is explicitly taken into
account has been presented in
Ref.\ \onlinecite{jjp_disk} and  also in Ref.\ \onlinecite{newpeeters2}. 
Part of the results presented here have already  appeared in Ref.\ 
\onlinecite{jjp_disk}.

\begin{figure}
\centerline {\epsfxsize=10cm \epsfbox{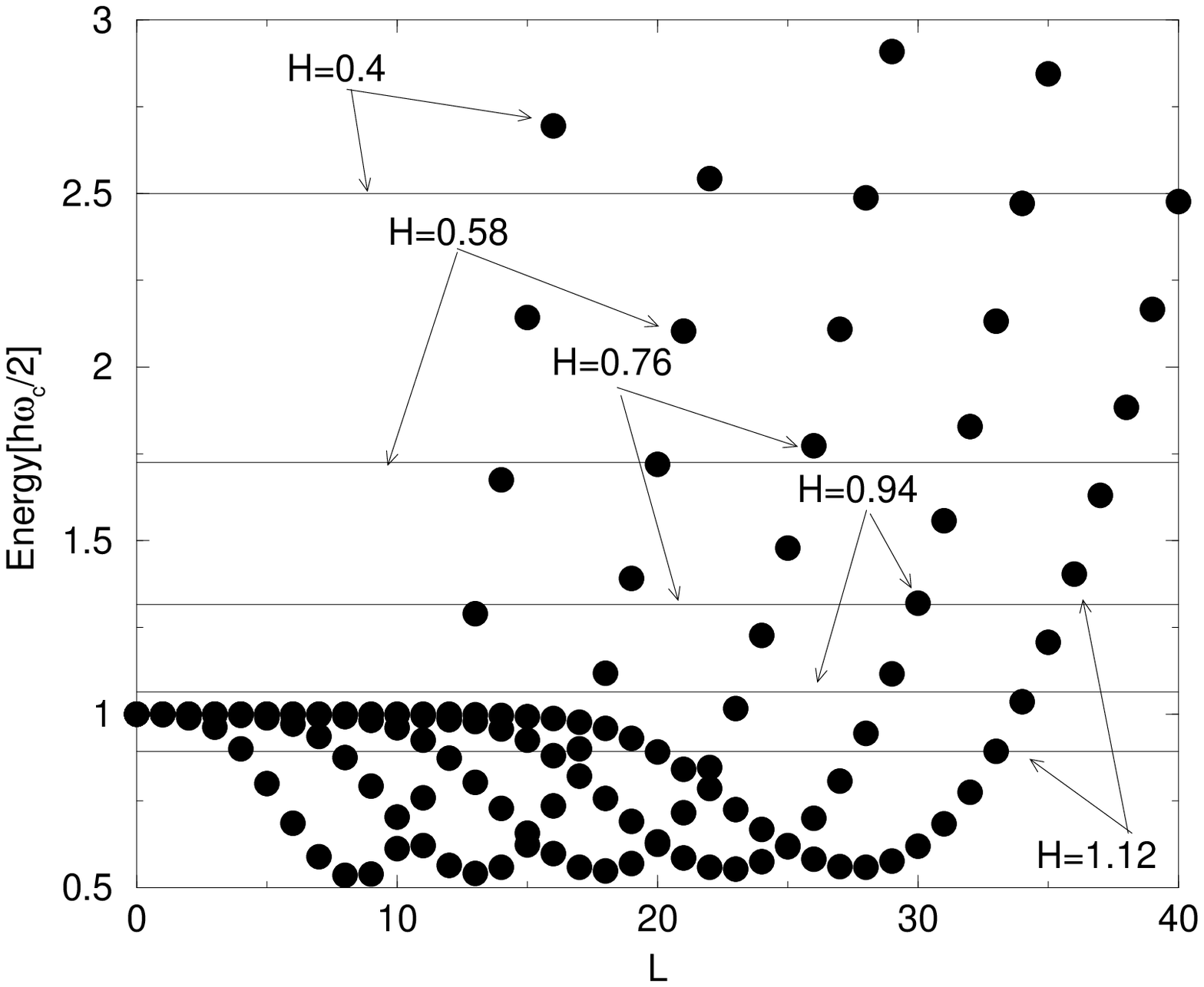}}
\caption{Band structure and chemical potential ($-\alpha$) of the Cooper
pairs at different values of $H$ (in units of $H_{\rm c2}$) for a 
disk of radius 
$R=8\xi$ and $\kappa=\infty$. Notice that,  for fields above $H_{\rm c2}$,
the lowest Landau level can lie below the chemical potential 
only close to the surface of the disk. This
gives rise to the formation of a giant vortex with an $L$ around the
bottom of the band.}
\label{bands_nij}
\end{figure}

We start from the traditional Ginzburg-Landau functional for the Gibbs
free energy of the superconducting state
\begin{eqnarray}
G_{\rm s}&=&G_{\rm n}+\int d{\bf r} \left[ \alpha |\Psi({\bf r})|^2 + 
\frac{\beta}{2}|\Psi({\bf r})|^4 + \right.\nonumber \\
&&\left.\frac{1}{2m^*}\Psi^*({\bf r})\left(-i\hbar{\bf\nabla} - 
\frac{e^*}{c}{\bf A({\bf r})}\right)^2\Psi({\bf r}) +
\frac{[h({\bf r})-H]^2}{8\pi} \right],
\label{G-L}
\end{eqnarray}
where $G_n$ is the Gibbs free energy of the normal state and $[-i
\hbar \nabla - e^* {\bf A}({\bf r})/c]^2/2m^*$ is the kinetic energy
operator for Cooper pairs of charge $e^*=2e$ and mass $m^*=2m$ in a
vector potential ${\bf A}({\bf r})$ which is associated with the magnetic
induction $h({\bf r})$.  The parameters $\alpha$ and $\beta$ have the
usual meaning\cite{Tinkham}. 
Before proceeding any further we must stress a not fully appreciated
fact: Even for small values of $\kappa$ ($\approx 1$),
the magnetic induction is weakly varying in space down to fairly low fields 
($H\approx 0.5H_{\rm c2}$)\cite{Brandt}. 
Thus, it is a very good approximation to
consider a uniform magnetic induction [$h({\bf r})=B$] down to 
$H\approx 0.5H_{\rm c2}$ and to expand the order parameter 
in the lowest Landau level (LLL):
\begin{equation}
\Psi({\bf r})=\sum_{L=0}^{\infty} C_L \frac{1}{\ell\sqrt{2\pi}}
e^{-iL\theta}\Phi_L(r).
\label{expansion}
\end{equation}
In this expansion $C_L \equiv |C_L|e^{i\phi_L}$ are complex coefficients and 
$\frac{1}{\ell\sqrt{2\pi}}e^{-iL\theta}\Phi_L(r)$ are normalized
eigenfunctions of the kinetic energy operator in 
Eq.\ (\ref{G-L}) plus the boundary conditions of zero current through the
surface of the disk\cite{surfaceold}. (We only consider disk 
thicknesses smaller 
than the coherence length so that the order parameter can be taken constant
in the direction of the field.) Strictly speaking, these eigenfunctions are 
not LLL eigenstates. However, their radial part
$\Phi_L(r)$, which may be found numerically, is nodeless and coincides
with the radial function of the symmetric LLL eigenfunctions for small $L$.
Figure \ref{bands_nij} shows the Cooper pair band structure, i.e., the
corresponding eigenvalues $\epsilon_L$ at different values of the external 
field for a disk of radius $R=8\xi$ and $\kappa=\infty$ (this implies $H=B$).
The horizontal lines represent $-\alpha$ which can be thought of as the
chemical potential for the Cooper pairs.
The formation of the dispersionless LLL in the
center of the disk  as the magnetic length
$\ell=\sqrt{e^*\hbar/cB}$ becomes smaller than $R$ is clearly visible. 
The bending of the band close to the surface is a consequence
of the boundary condition and allows the nucleation of surface
superconductivity (or the formation of a giant vortex)
even when the  bulk remains in the normal state ($H> H_{\rm c2}$).

The expansion (\ref{expansion}) captures both the simplicity of 
the macrovortex (above $H_{\rm c2}$) when only one $C_L$ is expected to be 
different from zero and the full complexity of 
the order parameter (below $H_{\rm c2}$) when several components or
harmonics must participate. 
Direct substitution of such an expansion into Eq.\ \ref{G-L} 
and subsequent numerical minimization of the resulting expression is a 
daunting task bound to fail due to the large number of unknown variables 
involved. Instead, it is preferable to consider expansions 
in restricted sets $\{L_1,L_2,\dots,L_N\}$ of few $N$ components where 
$L_1<L_2<\dots<L_N$. The difference between the Gibbs free energies of
the normal and superconducting phases takes the following form for each set:
\begin{eqnarray}
G_{\rm s}-G_{\rm n}&=&\sum_{i=1}^N \alpha[1-B\epsilon_{L_i}(B)] |C_{L_i}|^2  +
\frac{1}{4}\alpha^2\kappa^2 B R^2 \times \nonumber \\
&&\left[ \sum_{i=1}^N I_{L_i}(B) |C_{L_i}|^4+
\sum_{j>i=1}^N 4 I_{L_iL_j}(B)|C_{L_i}|^2|C_{L_j}|^2 \right.+ \nonumber \\
&&\sum_{k>j>i=1}^N 4 \delta_{L_i+L_k,2L_j}
\cos(\phi_{L_i}+\phi_{L_k}-2\phi_{L_j})  \nonumber \\
&&I_{L_iL_jL_k}(B)|C_{L_i}||C_{L_j}|^2|C_{L_k}| + \nonumber \\
&& \sum_{l>k>j>i=1}^N 8 \delta_{L_i+L_l,L_j+L_k}
\cos(\phi_{L_i}+\phi_{L_l}-\phi_{L_j}-\phi_{L_k}) \nonumber \\
&&\left.I_{L_iL_jL_kL_l}(B)|C_{L_i}||C_{L_j}||C_{L_k}||C_{L_l}| \right]
+(B-H)^2,
\label{LLL}
\end{eqnarray}
where $G_{\rm s}-G_{\rm n}$ and $\alpha$ are given 
in units of $H_{\rm c2}^2V/8\pi$ ($V$ is the volume 
of the disk), $\epsilon_L(B)$ is given in units of the lowest Landau level
energy $\hbar\omega_c/2$ (with $\omega_c=e^*B/m^*c$), $R$ is 
in units of $\xi$, and $B$ and $H$ are given
in units of $H_{\rm c2}$. The terms proportional to $\alpha$ contain
the condensation and kinetic energy of the Cooper pairs. All the other terms, 
which are  proportional to $\alpha^2$, account for  the
``interaction'' between Cooper pairs.
There appear four types of these terms: (i) those proportional to
$I_L(B) \equiv \int dr\:r\: \Phi_{L}^4$,
reflecting the interaction between Cooper pairs occupying the same
quantum state $L$, (ii) those proportional to  $I_{L_iL_j}(B)\equiv 
\int dr\:r\: \Phi_{L_i}^2 \Phi_{L_j}^2$,
reflecting the interaction between Cooper pairs occupying different
quantum states, and (iii) the ones proportional to  $I_{L_iL_jL_k}(B)\equiv 
\int dr\:r\: \Phi_{L_i}\Phi_{L_j}^2 \Phi_{L_k}$
and proportional to  $I_{L_iL_jL_kL_l}(B)\equiv
\int dr\:r\: \Phi_{L_i}\Phi_{L_j} \Phi_{L_k}\Phi_{L_l}$  which, along
with the phases $\phi_L$, are responsible for the spatial correlation between 
vortices and the detailed structure of the DVS (see below). 
The non-linear dependence on $B$ of these integrals 
[as well as that of $\epsilon_L(B)$] comes from the existence of the 
disk surface.

In order to find the minimum Gibbs free energy for a given set we have to
minimize with respect to the moduli $|C_{L_1}|,\dots,|C_{L_N}|$, the phases 
$\phi_{L_1},\dots,\phi_{L_N}$ of the coefficients, and with respect to $B$. 
The minimum-energy set of components is picked up at the end. The
advantage of doing this selective minimization resides in our
expectation that a small number of components will suffice to describe
the order parameter for disks with radii of few coherent lengths.
As illustrative and relevant examples
we consider in the detail the solutions with one and two components. 
For $N=1$ the energy functional
is invariant with respect to the phase of the only coefficient so 
one can minimize analytically with respect to $|C_L|^2$ to obtain
\begin{equation}
G_{\rm s}-G_{\rm n}=-\frac{[1-B\epsilon_L]^2}{\kappa^2 B R^2 I_L}+(B-H)^2,
\label{1c}
\end{equation}
where we have dropped the implicit $B$-dependences.
Finally, the minimal value of $B$ and the minimum Gibbs free energy 
for each $L$ must be found numerically. 
The 2-component  solutions $\{L_1,L_2\}$
can be dealt with in a similar way. The energy functional
is invariant with respect to the phases so one can minimize analytically with 
respect to $|C_{L_1}|^2$ and $|C_{L_2}|^2$ to obtain
\begin{equation}
G_{\rm s}-G_{\rm n}=-\frac{[1-B\epsilon_{L_1}]^2 I_{L_2}+
[1-B\epsilon_{L_2}]^2
I_{L_1}-4[1-B\epsilon_{L_1}][1-B\epsilon_{L_2}]I_{L_1L_2}}
{\kappa^2 B R^2 [I_{L_1}I_{L_2}-4I_{L_1L_2}]}+(B-H)^2
\label{2c}
\end{equation}
and numerically with respect to $B$ to obtain the minimum energy for the
given pair of components. The superconducting condensate
looks generically like an $(L_2-L_1)$-vortex ring. It
is important to notice that $\alpha$ disappears from the final
expressions (\ref{1c}) and (\ref{2c}) which leaves us with
$\kappa$ as the only adjustable parameter when comparing with 
experiments. (This is also true for the more complex cases 
discussed below). For 3-component solutions (two vortex rings)
the energy functional is still invariant with respect to all the three 
phases whenever $L_1+L_3\ne 2L_2$, and, once again, the minimization 
with respect to $|C_{L_1}|^2$, $|C_{L_2}|^2$, and $|C_{L_3}|^2$ can be
done analytically. However, if $L_1+L_3= 2L_2$, 
the two rings have the same number of vortices and their
relative angular positions come into play through the term depending 
on the phases. There is, however, an obvious choice for these phases:
$\phi_{L_1}=0,\phi_{L_2}=0,\phi_{L_3}=\pi$. This choice
gives a negative contribution to the free energy which
reflects a lock-in position between the vortex rings.  
In general, the minimum-energy
solutions for disks are expected to have strongly overlapping components 
which invalidates any perturbative treatment of the terms that contain the 
phases\cite{jjp_strip}. Moreover, unlike simpler
geometries\cite{jjp_strip}, there is no direct connection between
number of components in which we expand the order parameter 
and number of vortices. This prompts us to seek solutions through
numerical minimization with respect to the moduli and $B$ for the
3-component cases just mentioned, and, for $N>3$, with respect to the moduli, 
the phases, and $B$ whenever the terms involving phases are present. 
For the disk sizes like the ones used in the experiment of 
Ref.\ \onlinecite{Geim} 2 and 3-component solutions suffice 
to capture the relevant physics.

\begin{figure}
\centerline {\epsfxsize=10cm\epsfbox{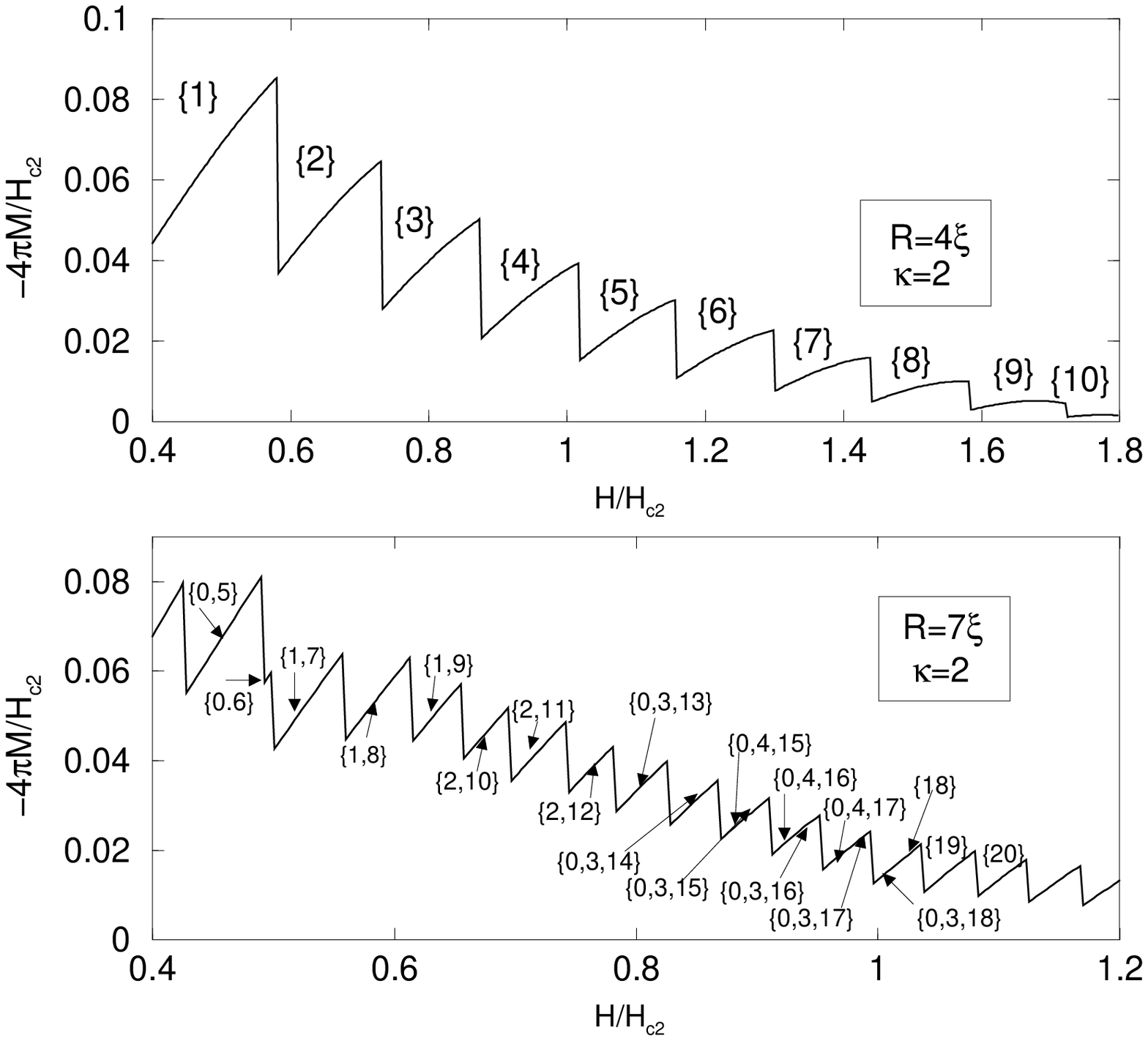}}
\caption{Magnetization as a function of $H$ for two disks of $\kappa=2$
and  radius (a) $R=4 \xi$ and (b)  $R=7 \xi$. The sets of
harmonics with significant weight corresponding to the
expansion of the order parameter at each magnetization step are also shown.}
\label{mag_nij}
\end{figure}

\begin{figure}
\centerline
{\epsfxsize=5cm\epsfysize=4.7cm\epsfbox{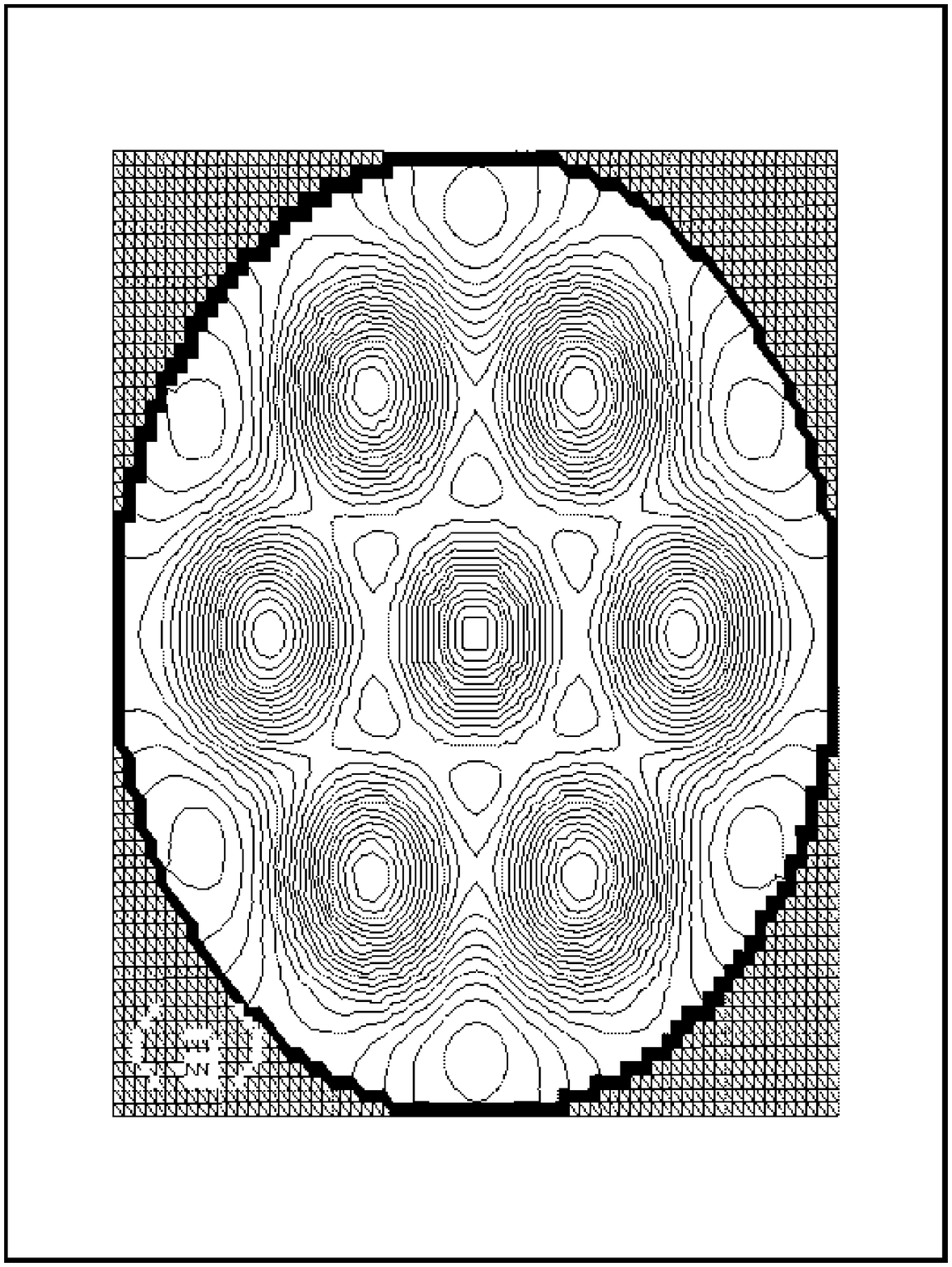}
\epsfxsize=5cm\epsfysize=4.7cm\epsfbox{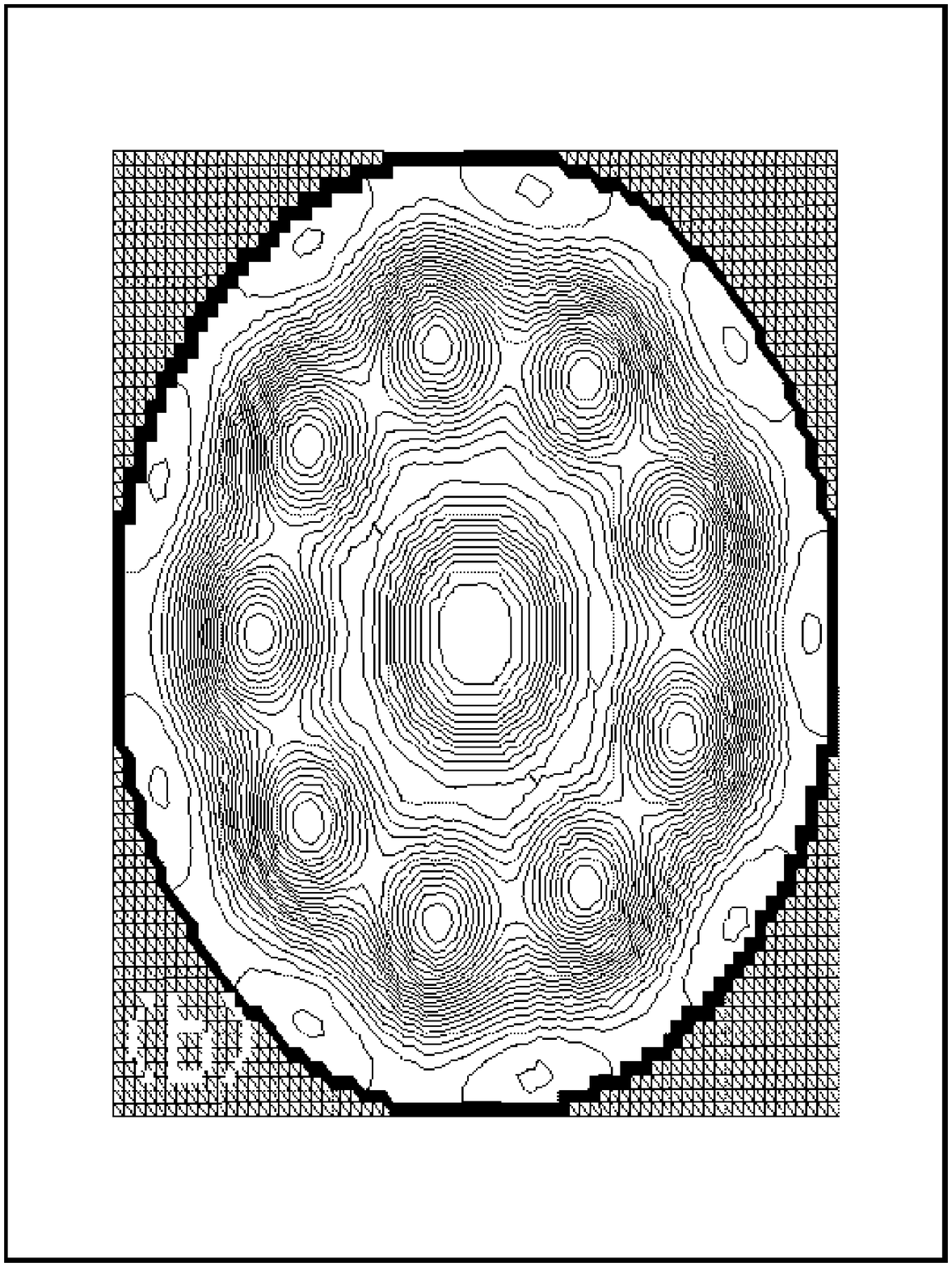}}
\centerline
{\epsfxsize=5cm\epsfysize=4.7cm\epsfbox{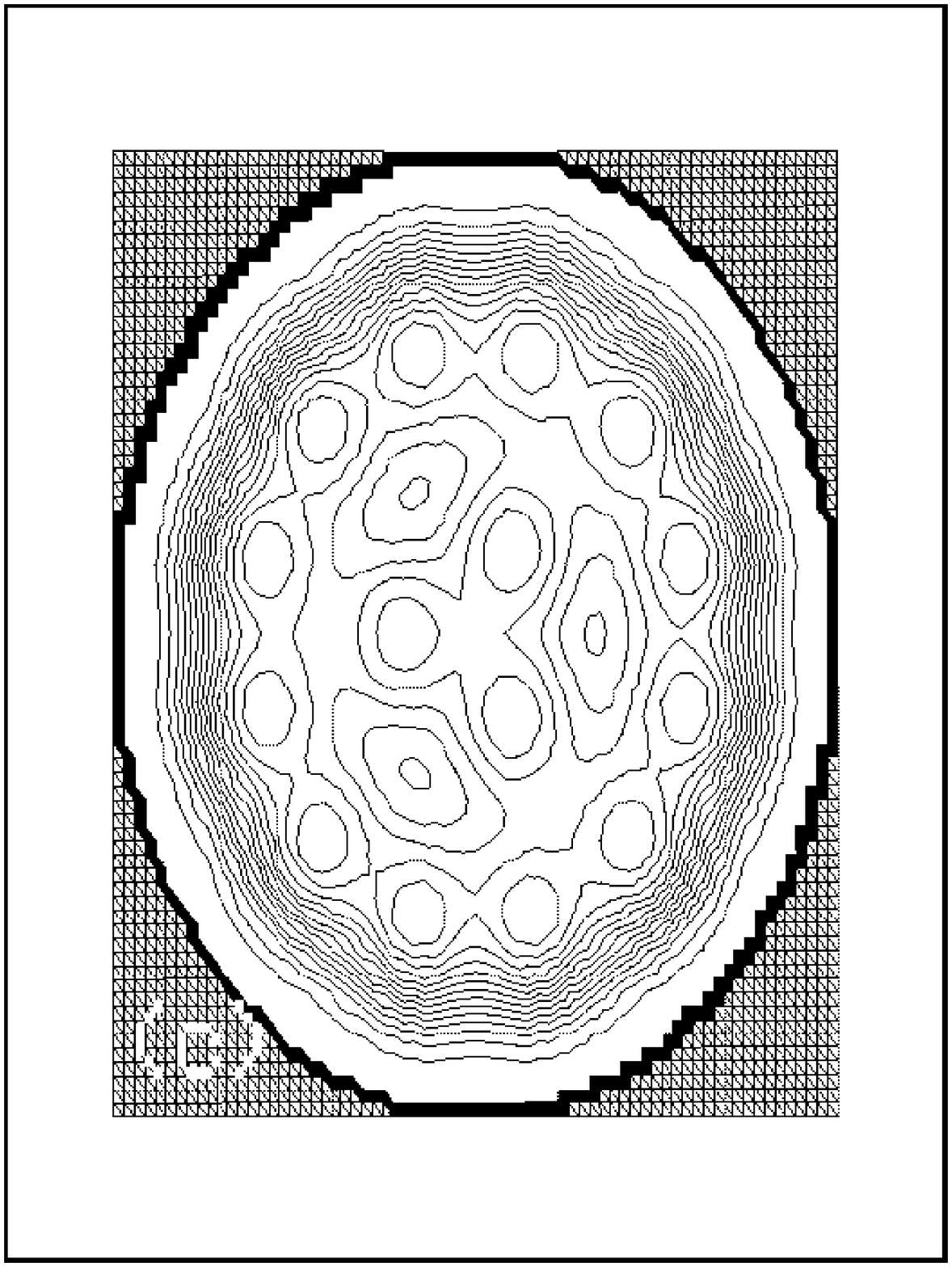}
\epsfxsize=5cm\epsfysize=4.7cm\epsfbox{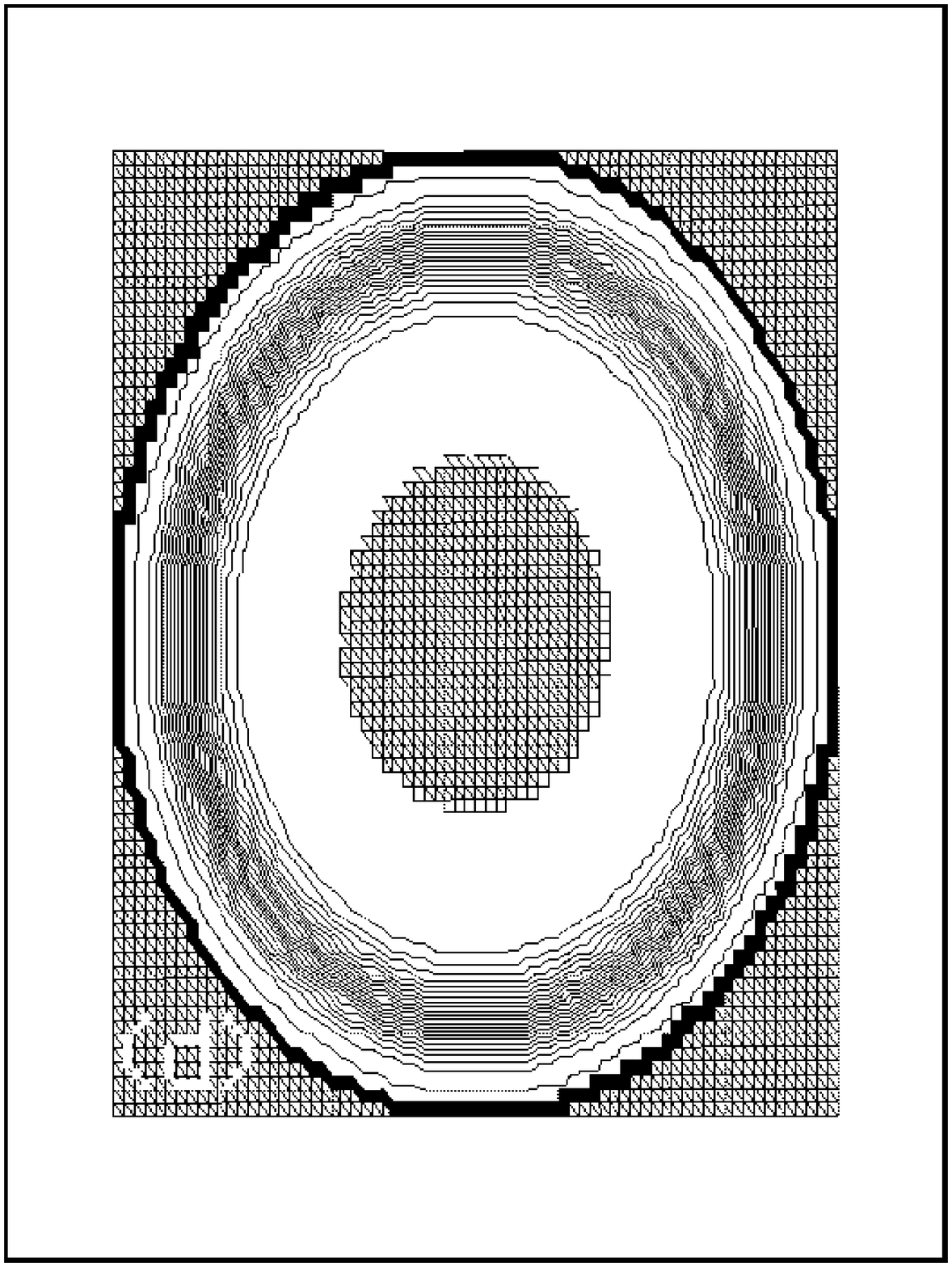}}
\caption{\protect Superconducting condensate in the DVS
for an $R=7 \xi$, $\kappa=2$ disk at
(a) $H=0.5H_{\rm c2}$, (b) $H=0.7H_{\rm c2}$, (c) $H=0.9H_{\rm c2}$, and (d)
$H=1.1H_{\rm c2}$. Below $H_{\rm c2}$ the density associated to the 
order parameter presents a vortex glass structure which disappears above
$H_{\rm c2}$ into a condensate of vortices or giant vortex. Notice the
appearance in (b) of two-fluxoid vortices even below $H_{\rm c2}$.}
\label{glass_nij}
\end{figure}

Figure \ref{mag_nij} shows the magnetization as a function of $H$ for
two disks with $\kappa=2$. The components or harmonics with significant
weight corresponding to the minimum-energy order parameter at each 
magnetization step are also shown. For the smaller disk of radius $R=4\xi$ we 
obtain a series of first-order transitions. 
Here all the minimal solutions correspond to giant
vortices which contain $L$ fluxoids. Whenever $L$ changes by one 
the magnetization presents a (non-quantized) jump whose magnitude 
evolves {\em monotonously} with $L$. For the larger $R=7\xi$ disk
$N>1$ solutions appear below $H_{\rm c2}$.
In this case, allowing more components in the expansion  of the order
parameter has a fundamental effect: It splits the giant
vortex into a complex vortex glass structure of many single-fluxoid
vortices [see Figs.\ \ref{glass_nij}(a) to (c)].
This glass structure reflects in the magnetization curves by
changing the regular evolution of the magnitude of the jumps into an
irregular one. Whenever a vortex is added or removed from the disk
the symmetry of the new vortex configuration changes
which, in turn, expels the field in a different way. 
The total number of vortices 
is always given by the largest $L_N$ which does not depend on
the value of $L$ of the other harmonics: Only the internal arrangement 
of vortices does. There usually exist configuration
switches for a given number of vortices (i.e., for a given $L_N$), 
but these changes {\em do not} 
reflect in the magnetization (see Fig\ \ref{mag_nij}). 
On top of the first-order transitions the overall 
slope in the magnetization clearly changes at $H_{\rm c2}$, i.e.,  at
the transition between the giant vortex and the vortex glass structures.
This transition is reminiscent of the second-order transition at
$H_{\rm c2}$ for bulk samples where $M$ vanishes. 

\begin{figure}
\centerline{\epsfxsize=14cm \epsfysize=8cm \epsfbox{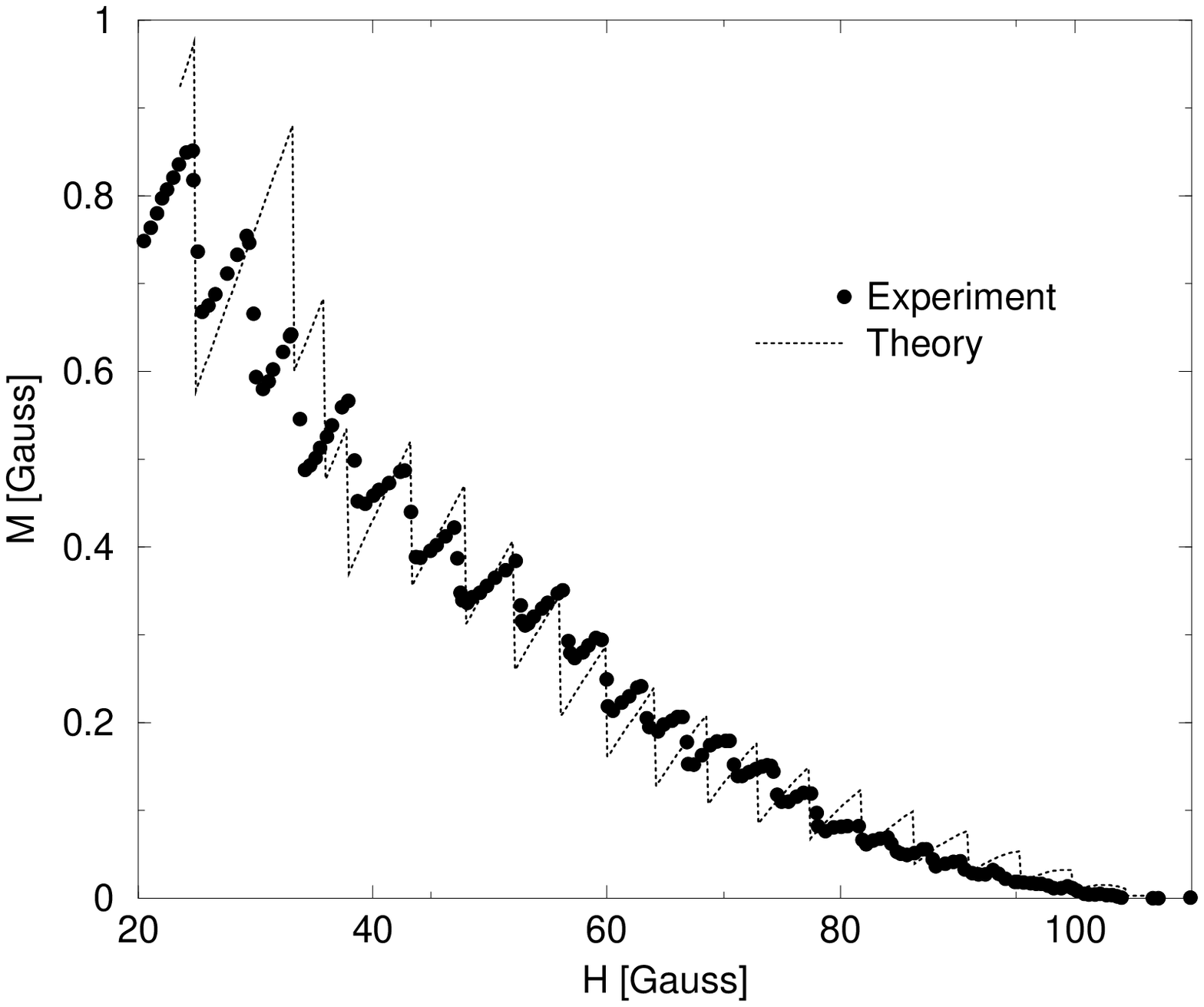}}
\caption{\protect Comparison between experimental data (reproduced from
Refs.\ [21]) and theory using $R=5.25\xi$ and $\kappa=1.2$. Similar
considerations as in Refs.\ [21] have been followed 
for the adjustment of the theoretical curve.}
\label{exp_nij}
\end{figure}

Finally, we point out that the magnetization measured by
Geim et al. in Ref.\ \onlinecite{Geim} presents features that are in
good agreement with our results despite
of the fact that the disks are made out of a strong type-I material as
Al. It is well known that the bulk value of $\kappa$ can increase in the plate 
geometry\cite{Dolan}.  To compare with the experiment, 
we simulate this fact by using a higher value of $\kappa$ than the nominal one. 
In Fig.\ \ref{exp_nij} we show the data for a disk of nominal
radius $R= 5\xi$ and thickness $d=0.6 \xi$. We have obtained a reasonable good
agreement in the number and magnitude of the jumps, and overall shape 
of the magnetization using $R \approx 5\xi$  and $\kappa\approx 1$ 
(the dotted line is a good example). This is consistent
with an effective  penetration length longer than expected
and, possibly, with a coherence length shorter 
than the bulk nominal one.  Although, due to the smallness 
of the disk, it is difficult to point at a
second-order phase transition, the non-monotonous evolution of the
magnitude of the magnetization jumps is notorious over a large 
range of fields which, as
we have shown, is an indication of the formation of vortex glass structures. 
In our approximation the magnetic induction is uniform in space, but,
given the good agreement with the experimental curve,
this does not seem to be an important restriction.

\section{Conclusions}
After reviewing some of the state-of-the-art experiments in mesoscopic
superconductivity we have shown how to calculate the 
dense vortex matter structure and associated magnetization
for type-II superconducting mesoscopic disks. We have found that the 
magnetization exhibits generically a first-order phase transition
whenever the number of vortices changes by one with $H$. It also
presents two well-defined regimes:  A {\em non-monotonous}
evolution of the magnitude of the
jumps signals the presence of a vortex glass structure which is
separated by a second-order phase
transition at $H_{\rm c2}$ from a condensed state of vortices (giant
vortex) where the magnitude of the jumps changes monotonously.
We have compared our results with the Hall magnetometry measurements
in Ref.\ \onlinecite{Geim} and
claimed that the magnetization exhibits clear
traces of the presence of these vortex glass states.          

\section{Acknowledgements}
The author acknowledges enlightening discussions with Andre Geim,
Francois Peeters, Carlos Tejedor and Joaqu\'{\i}n Fern\'andez-Rossier.
This work has been funded by NSF Grant DMR-9503814 and MEC of Spain under
contract No. PB96-0085.

\begin{references}
\bibitem{note} Notice that $H_{\rm c2}$ for a disk of an intrinsic
type-II material is larger than that for the same material in bulk 
since the surface superconductivity  carries a supercurrent and,
consequently, a magnetization. For simplicity's sake we are not making
any distinction here.
\end{references}
\end{document}